\documentclass[a4paper, twoside, reqno, 12pt, dvips]{amsart}
\RequirePackage{fix-cm} % Fix LaTeX2e bugs.

\usepackage{fixltx2e}     % Fix LaTeX2e bugs.

\usepackage{harvard}

\usepackage{eucal}
\usepackage{xspace}
\usepackage{amsgen}
\usepackage{amssymb}
\usepackage{amsmath}
\usepackage{amsfonts}
\usepackage{MnSymbol}
\usepackage[nice]{nicefrac}
\usepackage{mathrsfs, dsfont}

\usepackage{a4wide}

\usepackage{indentfirst}
\usepackage{graphicx}
\usepackage{psfrag}

\usepackage[usenames, dvipsnames, pdf]{pstricks}
\usepackage{epsfig}
\usepackage{pst-grad} % For gradients
\usepackage{pst-plot} % For axes

\usepackage{subfigure}

\usepackage[french,english]{babel}
\usepackage[latin1]{inputenc}

\usepackage{color}
\definecolor{oneblue}{rgb}{0,0.0,0.75}
\usepackage[colorlinks,
            urlcolor=oneblue,
            linkcolor=oneblue,
            citecolor=oneblue,
            bookmarksopen=false,
            pagebackref]{hyperref}

\numberwithin{equation}{section}

\newcommand{\R}{\mathbb{R}}
\newcommand{\U}{\mathbf{U}}
\newcommand{\V}{\mathbf{V}}
\newcommand{\ud}{\mathrm{d}}
\newcommand{\ui}{\mathrm{i}}
\newcommand{\ue}{\mathrm{e}}
\newcommand{\Ls}{\mathsf{L}}
\newcommand{\N}{\mathcal{N}}

\renewcommand{\v}{\mathbf{v}}
\newcommand{\eps}{\varepsilon}
\renewcommand{\O}{\mathcal{O}}
\renewcommand{\L}{\mathcal{L}}
\renewcommand{\H}{\mathcal{H}}

\newcommand{\RE}{\mathrm{Re}}

\newcommand{\p}{\partial}

\newcommand{\sech}{\mathrm{sech}}
\newcommand{\cn}{\mathop{\mathrm{cn}}}

\newcommand{\pd}[2]{\frac{\partial\, #1}{\partial\/ #2}}

\begin{document}

\title[Camassa--Holm equations and vortexons]%
{Camassa--Holm equations and vortexons for axisymmetric pipe flows}

\author[F.~Fedele]{Francesco Fedele$^*$}
\address{School of Civil and Environmental Engineering \& School of Electrical and Computer Engineering, Georgia Institute of Technology, Atlanta, USA}
\email{fedele@gatech.edu}
\urladdr{http://savannah.gatech.edu/people/ffedele/Research/}
\thanks{$^*$ Corresponding author}

\author[D.~Dutykh]{Denys Dutykh}
\address{University College Dublin, School of Mathematical Sciences, Belfield, Dublin 4, Ireland \and LAMA, UMR 5127 CNRS, Universit\'e de Savoie, Campus Scientifique, 73376 Le Bourget-du-Lac Cedex, France}
\email{Denys.Dutykh@ucd.ie}
\urladdr{http://www.denys-dutykh.com/}

\begin{abstract}
In this paper, we study the nonlinear dynamics of an axisymmetric disturbance to the laminar state in non-rotating Poiseuille pipe flows. In particular, we show that the associated Navier--Stokes equations can be reduced to a set of coupled Camassa--Holm type equations. These support inviscid and smooth localized travelling waves, which are numerically computed using the Petviashvili method. In physical space they correspond to localized toroidal vortices that concentrate near the pipe boundaries (wall vortexons) or wrap around the pipe axis (centre vortexons) in agreement with the analytical soliton solutions derived by \citeasnoun{Fedele2012b} for small and long-wave disturbances. Inviscid singular vortexons with discontinuous radial velocities are also numerically discovered as associated to special traveling waves with a wedge-type singularity, viz. peakons. Their existence is confirmed by an analytical solution of exponentially-shaped peakons that is obtained for the particular case of the uncoupled Camassa--Holm equations. The evolution of a perturbation is also investigated using an accurate Fourier-type spectral scheme. We observe that an initial vortical patch splits into a centre vortexon radiating vorticity in the form of wall vortexons. These can under go further splitting before viscosity dissipates them, leading to a slug of centre vortexons. The splitting process originates from a radial flux of azimuthal vorticity from the wall to the pipe axis in agreement with \citeasnoun{Eyink2008}. The inviscid and smooth vortexon is similar to the nonlinear neutral structures derived by \citeasnoun{Walton2011} and it may be a precursor to puffs and slugs observed at transition, since most likely it is unstable to non-axisymmetric disturbances.
\end{abstract}


\maketitle

\tableofcontents

\section{Introduction}

Transition to turbulence in non-rotating pipe flows is triggered by finite-amplitude perturbations \cite{Hof2003}, and the coherent structures observed at the transitional stage are in the form of localized patches known as puffs and slug structures \cite{Wygnanski1973,Wygnanski1975}. Puffs are spots of vorticity localized near the pipe axis surrounded by laminar flow, whereas slugs expand through the entire cross-section of the pipe while developing along the streamwise direction. Recent theoretical studies related slug flows to quasi--inviscid solutions of the Navier--Stokes (NS) equations. In particular, for non-axisymmetric flows \citeasnoun{Smith1982} revealed the existence of nonlinear neutral structures localized near the pipe axis (centre modes) that are unstable equilibrium states \cite{Walton2005}.  \citeasnoun{Walton2011} found the axisymmetric analogue of these inviscid traveling waves by studying the the nonlinear stability of impulsively started pipe flows to axisymmetric perturbations. Walton's modes are similar to the inviscid axisymmetric slug structures proposed by \citeasnoun{Smith1990}.

Recently \citeasnoun{Fedele2012b} investigated the dynamics of non-rotating axisymmetric pipe flows in terms of travelling waves of nonlinear soliton bearing equations. He showed that at high Reynolds numbers, the dynamics of small long-wave perturbations of the laminar flow obey a coupled system of nonlinear Korteweg--de Vries-type (KdV) equations. These set of equations generalize the one-component KdV model derived by \citeasnoun{Leibovich1968} (see also \citeasnoun{Leibovich1969}, \citeasnoun{Leibovich1984}) to study propagation of waves along the core of concentrated vortex flows (see also \citeasnoun{Benney1966}) and vortex breakdown \cite{Leibovich1984}. Fedele's coupled KdV equations support inviscid soliton and periodic wave solutions in the form of toroidal vortex tubes, hereafter referred to as \emph{vortexons}, which are similar to the inviscid nonlinear neutral centre modes found by \citeasnoun{Walton2011}. Fedele's vortical structures eventually slowly decay due to viscous dissipation on the time scale $t \sim \O(\RE^{6.25})$ \cite{Fedele2012b}. The vortexon, Walton's neutral mode and the inviscid axisymmetric slug proposed by \citeasnoun{Smith1990} are similar to the slugs of vorticity that have been observed in both experiments \cite{Wygnanski1973} and numerical simulations \cite{Willis2009}. As discussed by \citeasnoun{Walton2011}, these inviscid structures may play a role in pipe flow transition as precursors to puffs and slugs, since most likely they are unstable to non-axisymmetric disturbances \cite{Walton2005}.

In this paper, we extend Fedele's analysis and show that the axisymmetric NS equations for non-rotating pipe flows can be reduced to a set of soliton bearing equations of Camassa--Holm type \cite{Camassa1993,Dullin2003}. These support smooth and inviscid solitary waves that are numerically computed using the Petviashvili method (\cite{Petviashvili1976}, see also \cite{Pelinovsky2004,Lakoba2007,Yang2010}) confirming the validity of the theoretical solutions derived by \citeasnoun{Fedele2012b} for long-wave disturbances. Moreover, inviscid singular solitary waves in the form of peakons are numerically discovered, and the interpretation of the associated vortical structures are discussed. Finally,  the evolution of a perturbation to the laminar state is investigated within the framework of the proposed soliton equations.

\section{Camassa--Holm type equations for pipe flows}

Consider the axisymmetric flow of an incompressible fluid in a pipe of circular cross section of radius $R$ driven by an imposed uniform pressure gradient. Define a cylindrical coordinate system $(r, \theta, z)$ with the $z$-axis along the streamwise direction, and $(u,v,w)$ as the radial, azimuthal and streamwise velocity components. The time, radial and streamwise lengths as well as velocities are rescaled with $T$,$R$ and $U_0$ respectively. Here, $T = R/U_0$ is a convective time scale and $U_0$ is the maximum laminar flow velocity. The Stokes streamfunction $\psi$ of a perturbation $\bigl(u = -r^{-1}\p_z\psi, w = r^{-1}\p_r\psi\bigr)$ to the laminar base flow $W_0(r) = 1-r^2$ satisfies the nonlinear equation \cite{Itoh1977}
\begin{equation}\label{eq:fg}
  \p_t\Ls\psi + W_0\p_z\Ls\psi - \frac{1}{\RE}\Ls^2\psi = \N(\psi),
\end{equation}
where the nonlinear differential operator
\begin{equation*}
  \N(\psi) = -\frac{1}{r}\p_r\psi\p_z\Ls\psi + \frac{1}{r}\p_z\psi\p_r\Ls\psi - \frac{2}{r^2}\p_z\psi\Ls\psi,
\end{equation*}
the linear operator
\begin{equation*}%\label{eq:oper}
  \Ls = \L + \p_{zz}, \qquad \L = \p_{rr} - \frac{1}{r}\p_r = r\p_r\left(\frac{1}{r}\p_r\right),
\end{equation*}
and $\RE$ is the Reynolds number based on $U_0$ and $R$. The boundary conditions for (\ref{eq:fg}) reflect the boundedness of the flow at the centerline of the pipe and the no-slip condition at the wall, that is $\p_r\psi = \p_z\psi = 0$ at $r = 1$.

Drawing from \cite{Fedele2012b}, the solution of (\ref{eq:fg}) can be given in terms of a complete set of orthonormal basis $\{\phi_j(r)\}$ as
\begin{equation}\label{eq:ex1}
  \psi(r,z,t) = \sum_{j=1}^\infty \phi_j(r) B_j(z,t),
\end{equation}
where $B_j$ is the amplitude of the radial eigenfunctions $\phi_j$, which satisfy the Boundary Value Problem (BVP) \cite{Fedele2005,Fedele2012b}
\begin{equation}\label{eq:eig}
  \L^2\phi_j = -\lambda_j^2\L\phi_j,
\end{equation}
with $r^{-1}\phi_j$ and $r^{-1}\p_r\phi_j$ bounded at $r = +0$, and $\phi_j =\p_r\phi_j = 0$ at $r = 1$. Since $\phi_j$ satisfies the pipe flow boundary conditions \emph{a priori}, so does $\psi$ of (\ref{eq:ex1}). Note that the vorticity of the velocity field associated to the truncated expansion for $\psi$ is divergence-free. The positive eigenvalues $\lambda_j$ are the roots of $J_2(\lambda_j) = 0$, where $J_2(r)$ are the Bessel functions of first kind of second order (see \cite{Abramowitz1965}). The corresponding eigenfunctions
\begin{equation*}
  \phi_n = \frac{\sqrt{2}}{\lambda_n}\left[r^2 - \frac{r J_1(\lambda_n r)}{J_1(\lambda_n)}\right],
\end{equation*}
form a complete and orthonormal set with respect to the inner product 
\begin{equation*}\label{eq:scal}
  \left\langle \varphi_1, \varphi_2\right\rangle = -\int\limits_0^1\varphi_1\;\L\varphi_2\; r^{-1}\,\ud r = \int\limits_0^1\p_r\varphi_1\;\p_r\varphi_2 r^{-1}\,\ud r.
\end{equation*}
A Galerkin projection of (\ref{eq:fg}) onto the vector space $\mathcal{S}$ spanned by the first $N$ least stable modes $\{\phi_j\}_{j=1}^N$ yields a set of coupled Camassa--Holm (CH) type equations \cite{Camassa1993,Dullin2003,Dullin2004}
\begin{equation}\label{eq:CH}
  \p_t B_j + c_{jm}\p_z B_m + \beta_{jm}\p_{zzz}B_m + \alpha_{jm}\p_{zzt} B_m + N_{jnm}(B_n, B_m) + \frac{\lambda_j^2 B_j}{\RE} = 0,
\end{equation}
where $j = 1,\ldots,N$, the nonlinear operator
\begin{equation}\label{eq:N}
  N_{jnm}(B_n, B_m) = F_{jnm}B_n\p_z B_m + \\ G_{jnm}\p_z B_n\p_{zz}B_m + H_{jnm}B_n\p_{zzz}B_m,
\end{equation}
the coefficients $c_{jm}$, $\beta_{jm}$, $\alpha_{jm}$, $F_{jnm}$, $G_{jnm}$, $H_{jnm}$ are given in ~\ref{app:a} and summation over repeated indices $n$ and $m$ is implicitly assumed. A physical interpretation of the CH equations (\ref{eq:CH}) is as follows: the perturbation is given by a superposition of radial structures (the eigenmodes $\phi_j$) that nonlinearly interact while they are advected and dispersed by the laminar flow in the streamwise direction.

Note that CH type equations arise also as a regularized model of the 3-D NS equations \cite{Chen1999a,Domaradzki2001,Foias2001,Foias2002}, the so called Navier--Stokes-alpha model.

\section{Is there wave dispersion in axisymmetric Navier--Stokes flows?}

The Galerkin projection described above yields the dispersive CH type equations (\ref{eq:CH}) for the space-time evolution of the streamfunction $\psi$. The term $\partial_{txx}\psi$ arises also in the Benjamin--Bona--Mahony (BBM) equation \cite{bona}. It has the property to suppress dispersion, attenuating the dispersive effects induced by the KdV term $\partial_{xxx}\psi$. Indeed, consider the linear equation with both BBM and KdV dispersion
\begin{equation*}
  A_t - \alpha A_{xxt} + cA_{x} + \gamma A_{xxx} = 0.
\end{equation*}
The associated linear phase speed of a Fourier wave $\ue^{\ui(kx - \omega t)}$ is
\begin{equation*}
  C(k) = \frac{\omega}{k} = \frac{c - \gamma k^2}{1 + \alpha k^2},
\end{equation*}
and as $k\to\infty$, $C_0\to -\gamma/\alpha$. This implies that Fourier waves with large wavenumbers tend to travel at the same speed, that is dispersion is suppressed at high $k$'s, if $\alpha\neq 0$. As a result, self-steepening induced by nonlinearities can become dominant and blow-up is possible in finite time, or the two contrasting effects can balance each other leading to a peakon solution \cite{Dullin2001,Dullin2003}. The extreme case of dispersion suppression is when there is no dispersion, that is $C(k) = c$, as in the dispersionless CH equation, which also admits peakons  \cite{Camassa1993}. Clearly, if one adds a fifth--order dispersion term $A_{xxxxx}$, then $C(k)$ will grow as $k\to\infty$, and peakons do not exist since dispersion is too strong.

The CH/KdV dispersion is associated to a hidden `\emph{elastic energy}' that has no counterpart in axisymmetric Navier--Stokes flows, which are essentially two-dimensional (2-D) since vortex stretching is absent. To understand the physical origin of such wave dispersion, we consider the 2-D Euler equations for an inviscid fluid over the domain $\Omega$ in cartesian coordinates. The divergent-free velocity field is given by
\begin{equation*}
  \v = \left(-\pd{\psi}{y},\; \pd{\psi}{x}\right),
\end{equation*}
where $\psi$ is the streamfunction and the vorticity
\begin{equation*}
  \omega = \triangle\psi.
\end{equation*}
The equation of motion is
\begin{equation}\label{w}
  \pd{\omega}{t} = -\v\cdot\nabla\omega = -[\psi,\; \omega],
\end{equation}
where the commutator
\begin{equation*}
  [f, \;g] = \pd{f}{x}\pd{g}{y} - \pd{f}{y}\pd{g}{x}
\end{equation*}
It will be useful to consider the Hamiltonian formulation of (\ref{w}). Following \citeasnoun{Morrison1998}, this is given by
\begin{equation}\label{dw}
  \pd{\omega}{t} = \left\{\omega, \H\right\},
\end{equation}
where
\begin{equation*}%\label{H}
  \H = \frac{1}{2}\int_{\Omega}\left|\nabla\psi\right|^{2}\;\mathrm{d}\Omega = -\frac{1}{2}\int_{\Omega}\omega\psi\; \mathrm{d}\Omega.
\end{equation*}
is the kinetic energy of the system and the non-canonical Lie-Poisson bracket is defined as
\begin{equation*}%\label{bracket}
\left\{F,\; G\right\} = \int_{\Omega}\omega\left[\frac{\delta F}{\delta\omega},\frac{\delta G}{\delta\omega}\right]\;\mathrm{d}\Omega,
\end{equation*}
where $\delta$ denotes variational derivative. Clearly, $\H$ is an invariant of motion becuase of the anti-symmetry of the Poisson bracket, i.e. $\left\{ F,G\right\} = -\left\{ G,F\right\}$. The Hamiltonian structure of (\ref{dw}) yields a physical interpretation of the fluid motion in terms of a deformation of a 2-D membrane. Indeed, the Hamiltonian $\H$ can be interpreted as the elastic energy of a membrane subject to tensional forces. The surface $\psi(x,y)$ represents the displacements of the deformated membrane and the vorticity $\omega$ is proportional to the mean curvature $\kappa$ of $\psi$. This changes according to (\ref{dw}), while the elastic energy $\H$ is kept invariant. As the curvature $\kappa$ evolves in space and time, viz. vorticity is swept around $\Omega$ and changes in time, the surface $\psi$ locally bends sharply if $\kappa$ increases, or flattens if $\kappa$ decreases. Since the velocity streamlines are the contours of $\psi$, this implies that the vortical flow intensifies (attenuates) in regions of high (low) curvature of $\psi$.

The wave dispersion associated to the `\emph{elastic energy}' $\H$ can be revealed if we express (\ref{w}) solely in terms of $\psi$, that is
\begin{equation*}
\partial_{t}\Delta\psi = -\partial_{y}\psi\partial_{x}\Delta\psi + \partial_{x}\psi\partial_{y}\Delta\psi.
\end{equation*}
Here, the left--hand side yields the terms $\partial_{txx}\psi$ and $\partial_{tyy}\psi$ that are typical of the CH equation. They indicate that as the vorticity changes in time, so does the curvature $\kappa$ of the surface $\psi$, which elastically deforms while the `\emph{energy}' is conserved. If the velocity field is given by the sum of a base flow and a perturbation, then KdV type dispersive terms $\partial_{xxx}\psi$ and $\partial_{yyy}\psi$ arise from the convection of the perturbation by the mean flow.

The Navier--Stokes--alpha model can be interpreted in a similar manner \cite{Foias2001}. This is given by
\begin{equation*}
\begin{array}{c}
\partial_{t}\V + \U\cdot\nabla\V + \nabla\U^{T}\cdot\V + \nabla p = \nu\Delta\V\\
\\
\nabla\cdot \U = 0,
\end{array}
\end{equation*}
and $\V = (1-\alpha^{2}\Delta)\U$. The typical Camassa--Holm terms arise from $\partial_{t}\Delta\U$. If $\U$ is the sum of a base flow and a perturbation, then KdV type dispersive terms arise as well.

\section{Long-wave limit and KdV vortexons}

As $\RE \to \infty$, \citeasnoun{Fedele2012b} showed that the nonlinear dynamics of a small long-wave perturbation $b_j = \eps B_j$, with $\eps \sim \O(\RE^{-2/5})$, can be reduced to that on the slow manifold of the laminar state spanned by the first few $N$ least stable modes, and higher damped modes are neglected. This is legitimate as long as the amplitudes $B_j$ remain small for all time and the non-resonant condition
\begin{equation}\label{eq:nonrescond}
 \lambda_{i_1}^2 + \lambda_{i_2}^2+\ldots \lambda_{i_k}^2 \neq \lambda_j^2 
\end{equation}
is satisfied for any permutation $\{i_1, i_2,\ldots, i_k\}$ of size $k\leq N$ drawn from the set $j=1,\ldots, N$ \cite{DelaLlave1997}. For the BVP of (\ref{eq:eig}) the relation (\ref{eq:nonrescond}) is verified numerically to hold up to $N \cong 10^4$. For time scales much less than $t \sim \O(\eps^{-2.5}) \cong \O(\RE^{6.25})$, the nonlinear dynamics of (\ref{eq:CH}) is primilary inviscid and obeys a set of coupled KdV equations \cite{Fedele2012b}
\begin{equation}\label{eq:KdV}
\p_{\tau}b_j + \tilde{\beta}_{jm}\p_{\xi\xi\xi}b_j + \tilde{F}_{jnm}b_n\p_\xi b_m = 0,
\end{equation}
defined on the stretched reference frame
\begin{equation*}%\label{eq:scaling}
  \xi = \eps^{1/2}(z - Vt), \qquad \tau = \eps^{3/2}t,
\end{equation*}
where the tensors $\tilde{\beta}_{jm}$, $\tilde{F}_{jnm}$ are given in \citeasnoun{Fedele2012b} and the celerity $V$ is, with good approximation, the average of the eigenvalues of $c_{jm}$. The nonlinear system (\ref{eq:KdV}) support analytical travelling waves (TW), for example,
\begin{equation}\label{eq:tw}
  b_j^{(tw)}(\xi, \tau) = k^2 x_j\left[-\frac{2 M^2 - 1}{3M^2} + \cn^2(k\xi)\right],
\end{equation}
where $\cn(\zeta)$ is the Jacobi elliptic function with modulus $0 \leq M \leq 1$, $k$ and $M$ are free parameters and $\{x_j\}\in\R^J$ is the intersection point of $J$ hyperconics $\Gamma_j$ given by
\begin{equation*}
  -12M^2\tilde{\beta}_{jj} x_j + \tilde{F}_{jnm}x_n x_m = 0, \qquad j = 1,\ldots, N.
\end{equation*}
For $M\to 1$, (\ref{eq:tw}) reduces to the family of localized sech-type solitary waves
\begin{equation}\label{eq:sech}
  b_j^{(s)}(\xi, \tau) = -\frac{1}{3} k^2 x_j + k^2 x_j\sech^2 (k\xi).
\end{equation}
In physical space, (\ref{eq:tw}) and (\ref{eq:sech}) represent respectively localized and periodic toroidal vortices, which travel slightly slower than the maximum laminar flow speed $U_0$, \emph{viz}. $V \approx 0.77U_0$. For $N = 2$, the vortical structures are localized near the wall (wall vortexon, $x_1$ and $x_2$ have same sign) or wrap around the pipe axis (centre vortexon, $x_1$ and $x_2$ have opposite sign). They have a non-zero streamwise mean, but they radially average to zero to conserve mass flux through the pipe. Vortexons may be related to the inviscid neutral axisymmetric slug structures discovered by \citeasnoun{Walton2011} in unsteady pipe flows, which are similar to the centre modes proposed by \citeasnoun{Smith1990}.

In the following we will compute numerically TWs of the inviscid CH-type equations (\ref{eq:CH}) and discuss the vortical structure of the associated disturbances.

\section{Regular and Singular Vortexons}

Consider the inviscid three-component CH equations (\ref{eq:CH}) with $N = 3$, and an ansatz for the wave amplitudes of the form $B_j = q + F_j(z-ct)$, where $q$ is a free parameter and $c$ is the velocity of the TW. The associated nonlinear steady problem for $F_j$ (in the moving frame $z-ct$) is solved using the Petviashvili method \cite{Petviashvili1976}, see also \cite{Pelinovsky2004,Lakoba2007,Yang2010}. This numerical approach has been successfully applied to derive TWs of the spatial Dysthe equation \cite{Fedele2011} and the compact Zakharov equation for water waves \cite{Fedele2012}. To initialize the iterative process, the initial guess for the wave components $B_j$ is set equal to the analytical cnoidal TW of the uncoupled KdV equations associated to (\ref{eq:CH}), \emph{viz}. $c_{jm} \approx c_{jj}$, $F_{jnm}\approx F_{jjj}$, and $\alpha_{jm} = G_{jnm} = H_{jnm} = 0$. Then, a converged solution is numerically continued by varying the parameters $c$ or $q$. Note that the parameter that controls the strength of the nonlinearity in the truncated Camassa--Holm equations is the travelling wave amplitude.

\begin{figure}
  \centering
  \includegraphics[trim=0.3cm 1.6cm 1cm 1.3cm, clip=true, width=0.5\textwidth]{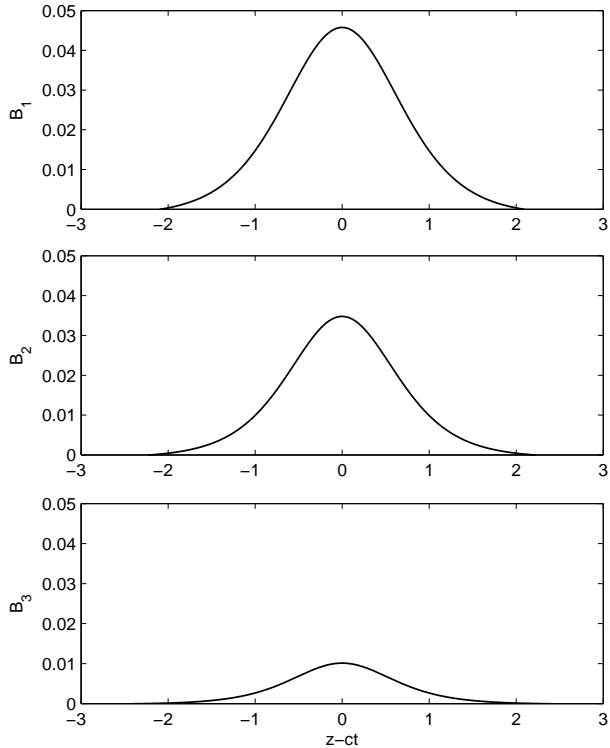}
  \caption{\small\em Inviscid regular wall vortexon: wave components $B_1$, $B_2$ and $B_3$ of the CH equations ($c = 0.65$, $q = 0$).}
  \label{fig:fig1}
\end{figure}

\begin{figure}
  \centering
  \includegraphics[trim=0.6cm 0.7cm 1cm 0.4cm, clip=true, width=0.55\textwidth]{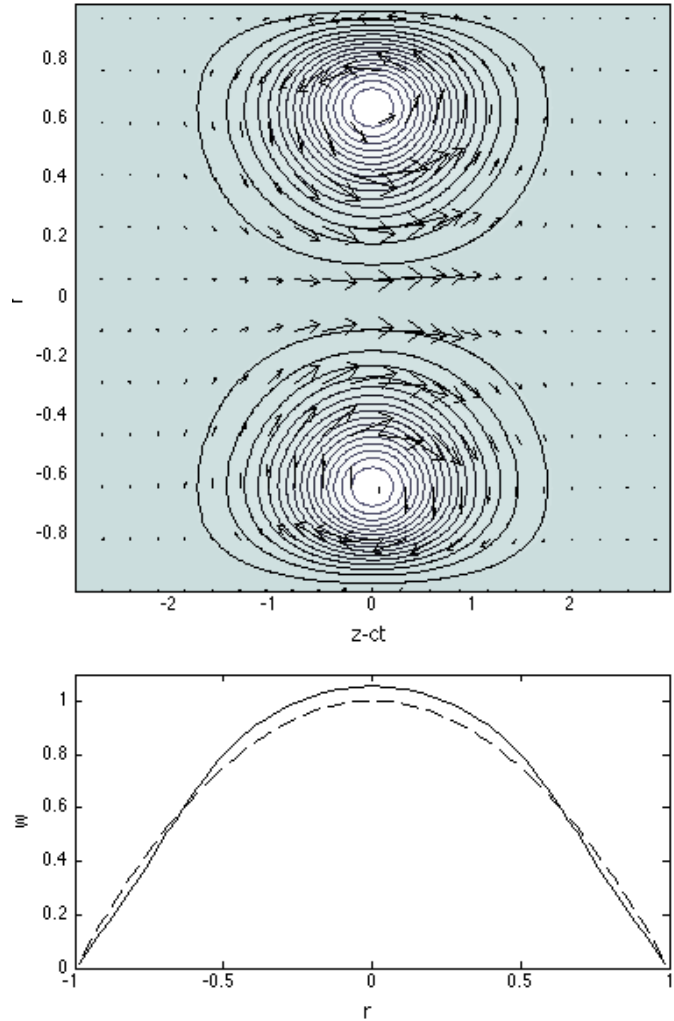}
  \caption{\small\em Inviscid regular wall vortexon: (top) streamlines of the three- component CH solution of Fig.~\ref{fig:fig1} and (bottom) velocity profiles of the perturbed (solid) and laminar (dash) flows ($c = 0.65$, $q = 0$).}
  \label{fig:fig2}
\end{figure}

The numerical basin of attraction of the Petviashvili scheme to localized TWs (solitons or solitary waves) is very sparse over the parameter space $(c,q)$. The generic topology of the flow structure associated to converged smooth TWs  is the same as that of the theoretical counterpart derived by \citeasnoun{Fedele2012b}: toroidal tubes of vorticity localized near the pipe boundaries (wall vortexons) or that wrap around the pipe axis (centre vortexons). In particular, wall vortexons are found in parameter window $c \sim [0.58, 0.66]$ and $q = 0$, however the Petviashvili scheme did not converge for $q>0$. For example, for $c = 0.65$ the wave components $B_j$ are shown in Figure~\ref{fig:fig1} and the streamlines of the associated flow perturbation are reported in the top panel of Figure~\ref{fig:fig2}. The perturbed flow (laminar plus vortexon) is shown in the bottom panel of the same Figure.  Note that wave components of higher modes have smaller amplitudes as an indication that their effects may vanish as $N$ increases, but a more systematic numerical study of this trend is required. %We believe that the regular vortexons of the truncated Camassa--Holm equations (\ref{eq:CH}) are approximations of exact invariant axisymmetric solutions of the Navier-Stokes equations. However, a rigorous proof of this statement is far beyond the scope of this work, and we just point out that the proper theoretical framework for such a proof is provided by slow-manifold type theorems and normal hyperbolicity in dynamical systems \cite{DelaLlave1997}. 

\begin{figure}
  \centering
  \includegraphics[trim=0.6cm 0.7cm 1cm 0.4cm, clip=true, width=0.55\textwidth]{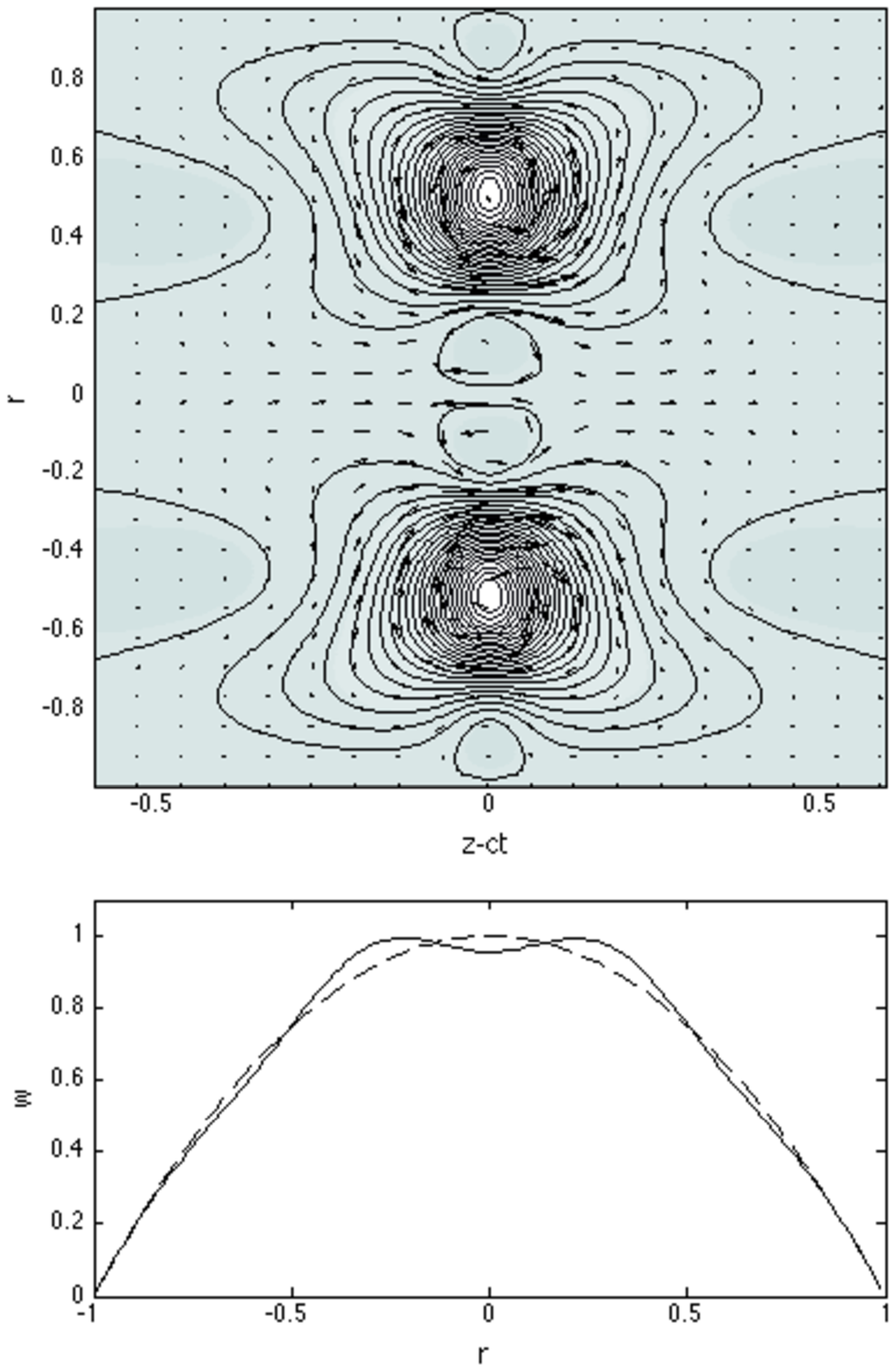}
  \caption{\small\em Inviscid regular wall vortexon: (top) streamlines of the three- component CH solution for $c = 0.78$, $q = 0$, and (bottom) velocity profiles of the perturbed (solid) and laminar (dash) flows.}
  \label{fig:fig3}
\end{figure}

\begin{figure}
  \centering
  \includegraphics[trim=0.5cm 0.7cm 1.1cm 0.4cm, clip=true, width=0.5\textwidth]{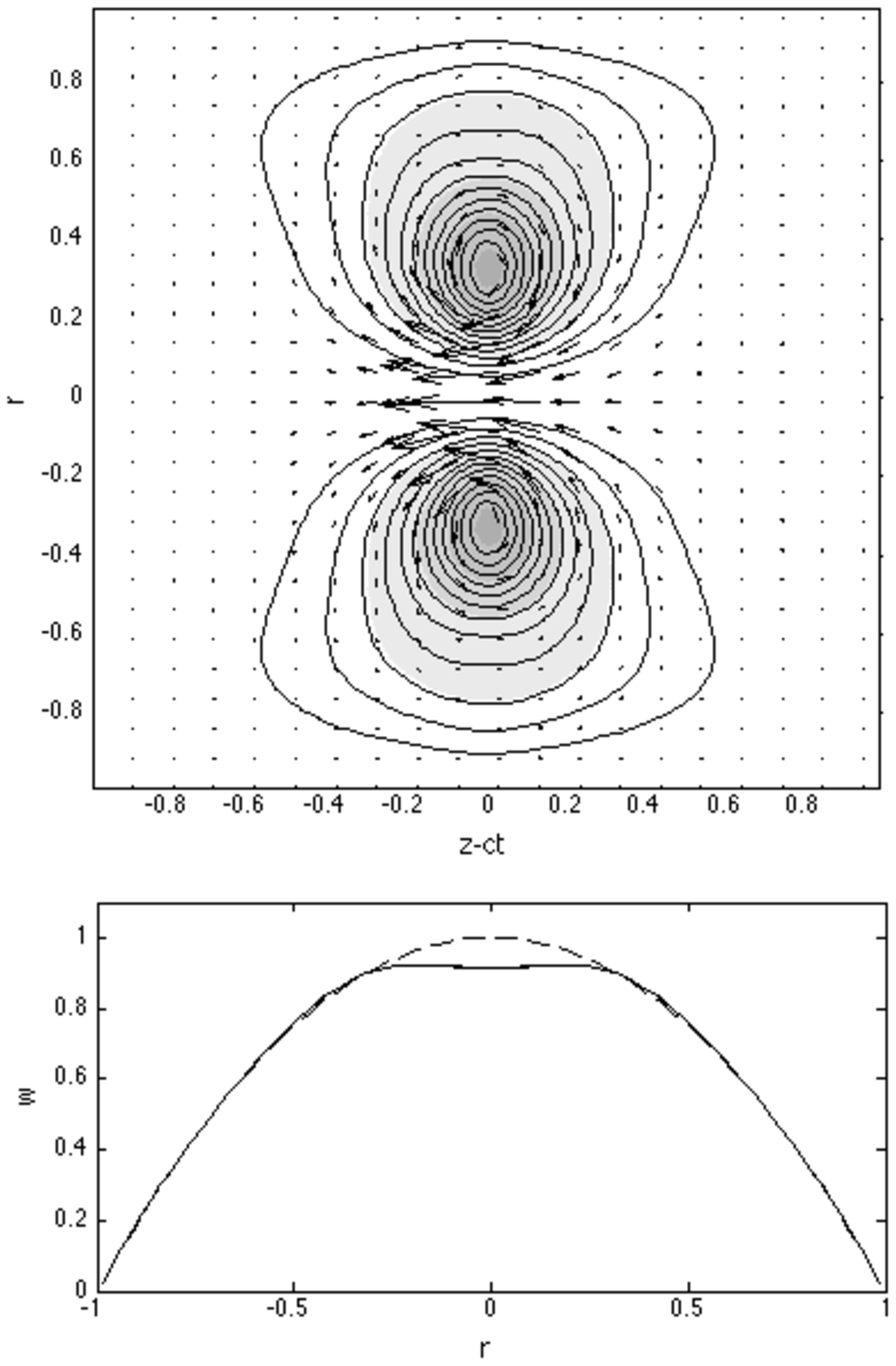}
  \caption{\small\em Inviscid regular centre vortexon: (top) streamlines of the three- component CH solution for $c = 0.86$, $q = 0$, and (bottom) velocity profiles of the perturbed (solid) and laminar (dash) flows.}
  \label{fig:fig4}
\end{figure}

\begin{figure}
  \centering
  \includegraphics[trim=0.3cm 1.15cm 1.3cm 1.2cm, clip=true, width=0.69\textwidth]{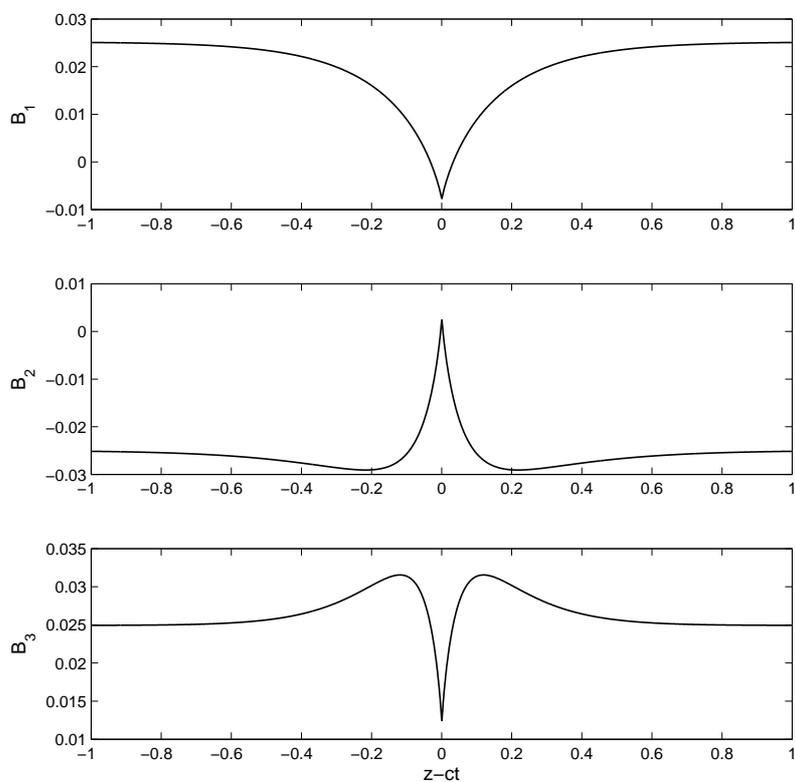}
  \caption{\small\em Inviscid singular vortexon: wave components $B_1$, $B_2$ and $B_3$ of the CH equations ($c = 0.90$, $q = 0.025$).}
  \label{fig:fig5}
\end{figure}

\begin{figure}
  \centering
  \includegraphics[trim=0.3cm 0.8cm 1.3cm 1cm, clip=true, width=0.59\textwidth]{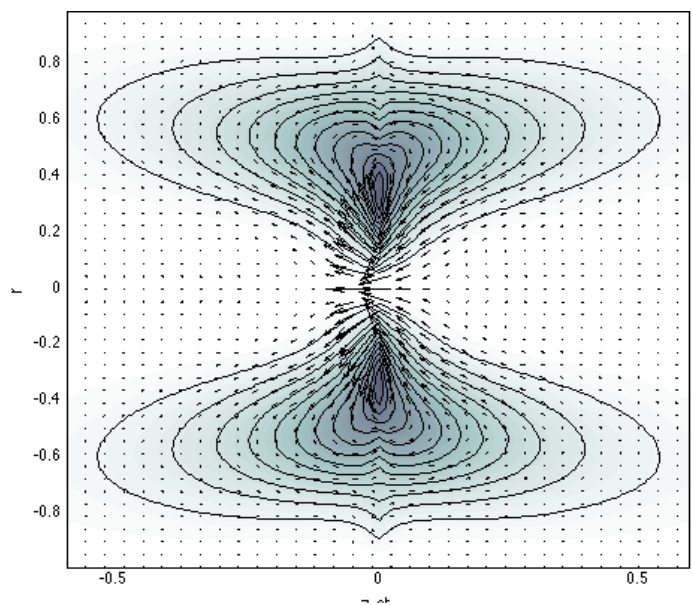}
  \caption{\small\em Inviscid singular vortexon: streamlines of the three- component CH solution of Fig.~\ref{fig:fig5} for $c = 0.90$, $q = 0.025$.}
  \label{fig:fig6}
\end{figure}

Convergence to inviscid wall vortexons also occurred in the range of $c \sim [0.762, 0.79]$ and $q = 0$ (it did not converge for $q>0$). For $c = 0.78$ the corresponding vortical structure is shown in Figure~\ref{fig:fig3}. Centre vortexons converged for $c \sim [0.82, 0.90]$ and $q = 0$ as depicted in Figure~\ref{fig:fig4} ($c = 0.86$). In this range of values of $c$ we note that as $q$ increases from zero, the smooth centre vortexon bifurcates to a traveling wave with a wedge-type singularity, \emph{viz}. peakon, as shown in Figure~\ref{fig:fig5} for $c = 0.90$, $q = 0.025$. In physical space the peakon corresponds to a localized vortical structure with discontinuous radial velocity $u$ across $z-ct=0$ (see Figure~\ref{fig:fig6}), but continuous streamwise velocity $w$ since the mass flux through the pipe is conserved. As a result, a sheet of azimuthal vorticity is advected at speed $c$.

\begin{figure}
  \centering
  \includegraphics[trim=0.3cm 0.5cm 1cm 0.5cm, clip=true, width=0.58\textwidth]{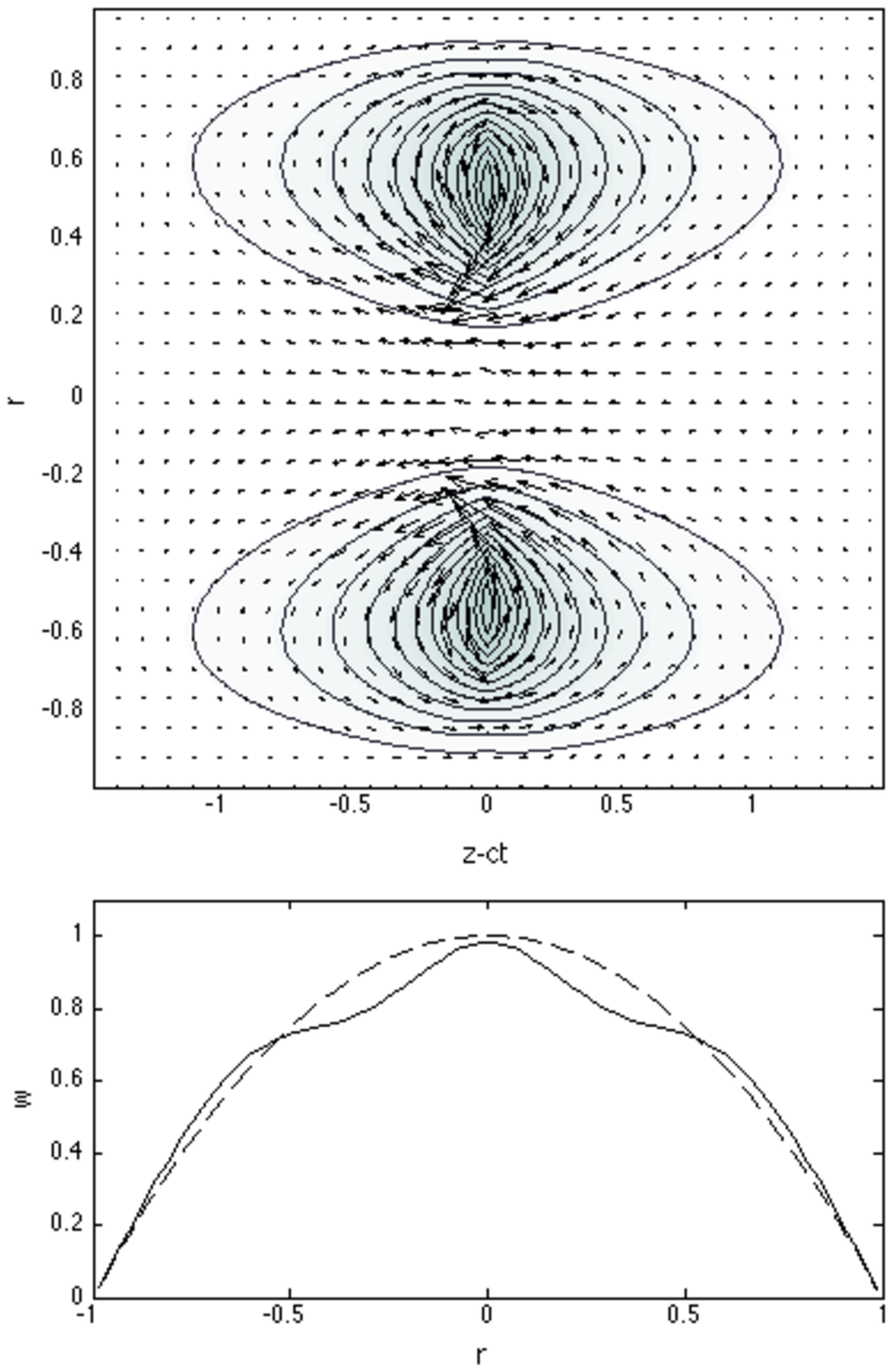}
  \caption{\small\em Inviscid singular vortexon: (top) streamlines of the three- component CH solution for $c = 0.70$, $q = 0$, and (bottom) velocity profiles of the perturbed (solid) and laminar (dash) flows.}
  \label{fig:fig7}
\end{figure}

\begin{figure}
  \centering
  \includegraphics[trim=0.5cm 1cm 2.2cm 1.2cm, clip=true, width=0.69\textwidth]{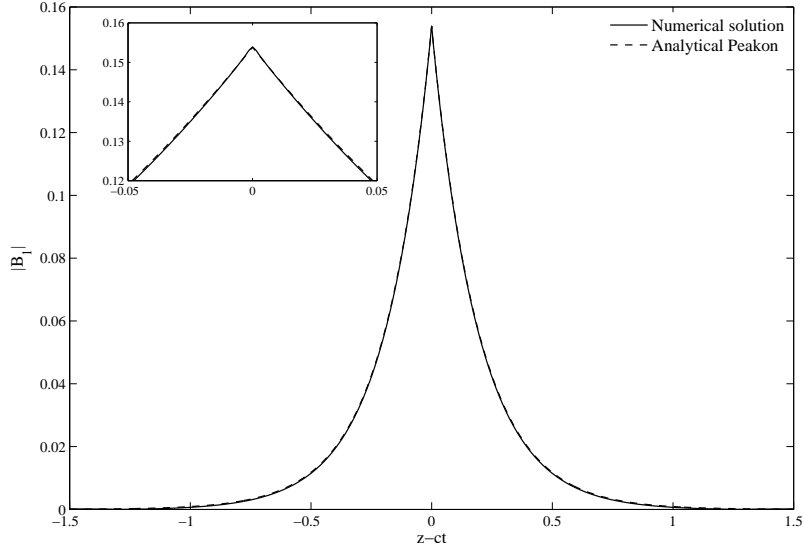}
  \caption{\small\em Analytical inviscid CH peakon (solid line) and numerical solution (dashed line) obtained by the Petviashili method (dimensionless velocity $c=V_1 \approx 0.63$).}
  \label{fig:fig8}
\end{figure}

\begin{figure}
  \centering
  \includegraphics[width=0.7\textwidth]{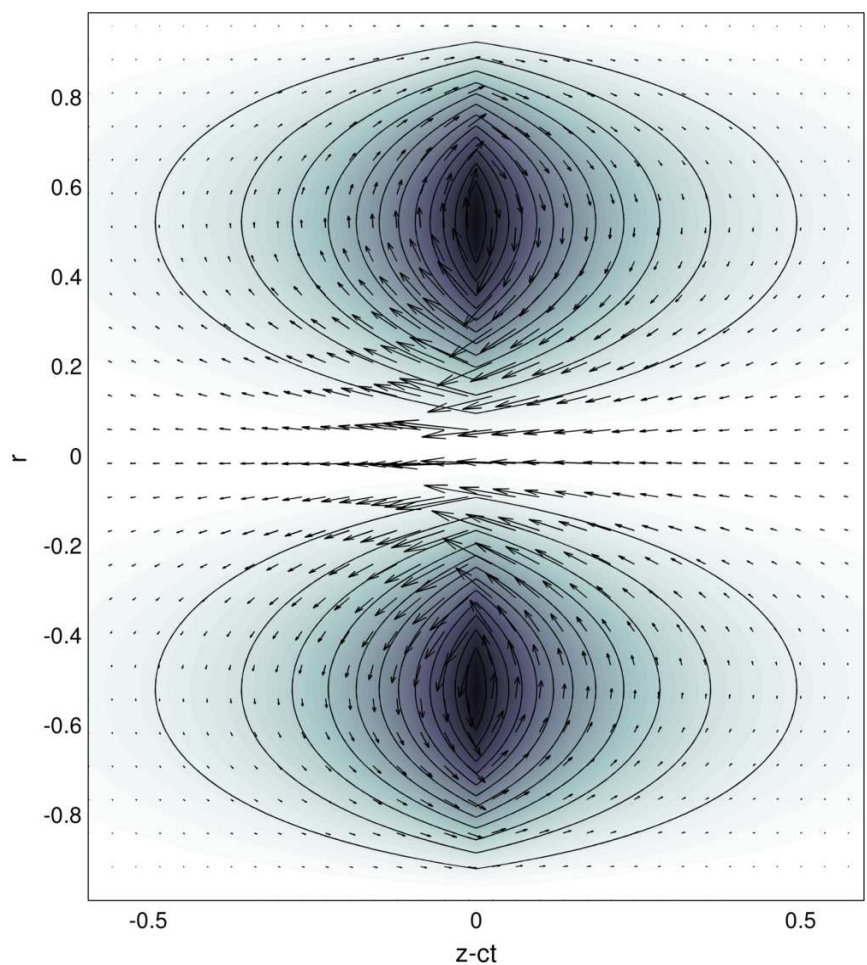}
  \caption{\small\em Inviscid singular vortexon associated to the peakon of Fig.~\ref{fig:fig8}: streamlines of the perturbation.}
  \label{fig:fig9}
\end{figure}

The Petviashvili method also converged to singular wall vortexons in the window $c \sim [0.69, 0.71]$ and only $q = 0$ as shown in Figure~\ref{fig:fig7} for the case of $c = 0.70$. The existence of singular vortexons is confirmed by an analytical solution of peakons obtained for the uncoupled version of the CH equations (\ref{eq:CH}), viz. 
\begin{equation}\label{eq:CH2}
  \p_t B_j + c_{jj}\p_z B_j + \beta_{jj}\p_{zzz}B_j + \alpha_{jj}\p_{zzt}B_j +\N_j(B_j) = 0,
\end{equation}
where
\begin{equation*}
  \N_j(B_j) = F_{jjj}B_j \p_z B_j + G_{jjj}\p_z B_j \p_{zz} B_j + H_{jjj} B_j\p_{zzz}B_j,
\end{equation*}
and here no implicit summation over repeated indices is assumed. Note that equation (\ref{eq:CH2}) is the dispersive Camassa--Holm equation with KdV dispersion, which admits peakon solutions \cite{Dullin2003}. These are given by (see ~\ref{app:b} for derivation)
\begin{equation}\label{eq:peak}
  B_j(z,t) = a_j \ue^{-\gamma_j\left\vert z-V_j t\right\vert},
\end{equation}
where
\begin{equation*}
  a_j = \frac{V_j\alpha_{jj} - \beta_{jj}}{H_{jjj}}, \qquad 
  V_j = \frac{c_{jj} + \beta_{jj} s_j^2}{1 + \alpha_{jj}s_j^2}, \qquad
  \gamma_j^2 = -\frac{F_{jjj}}{G_{jjj} + H_{jjj}}.
\end{equation*}
Note that the peakon arises as a special balance between the linear dispersion terms $\p_{zzz}B_j$, $\p_{zzt}B_j$ and their nonlinear counterpart $B_j\p_{zzz}B_j$ in (\ref{eq:CH2}). These three terms are interpreted in distributional sense because they give rise to derivatives of Dirac delta functions that must vanish by properly chosing the amplitude $a_j$, thus satisfying the differential equation (\ref{eq:CH2}) in the sense of distributions. The associated streamfunction $\psi_j^{(p)}$ is given by
\begin{equation*}%\label{eq:streamsol}
  \psi_j^{(p)}(r,z,t) = a_j \ue^{-\gamma_j\left\vert z-V_j t\right\vert}\phi_j(r).
\end{equation*}
For the least stable eigenmode $B_1$, Figure~\ref{fig:fig8} shows the remarkable agreement between the theoretical peakon (\ref{eq:peak}) and the associated numerical solution obtained via the Petviashvili method. The associated vortical structure (streamlines) is shown in Figure~\ref{fig:fig9} and it is similar to that of the numerical vortexons of Figures \ref{fig:fig6} and \ref{fig:fig7}.

Finally, note that viscous dissipation precludes the existence of peakons and slowly decaying smooth vortexons appear in the CH dynamics as discussed below.

\section{Vortexon slugs}

\begin{figure}
  \centering
  \includegraphics[trim=1.1cm 1.8cm 1.6cm 1.1cm, clip=true, width=0.89\textwidth]{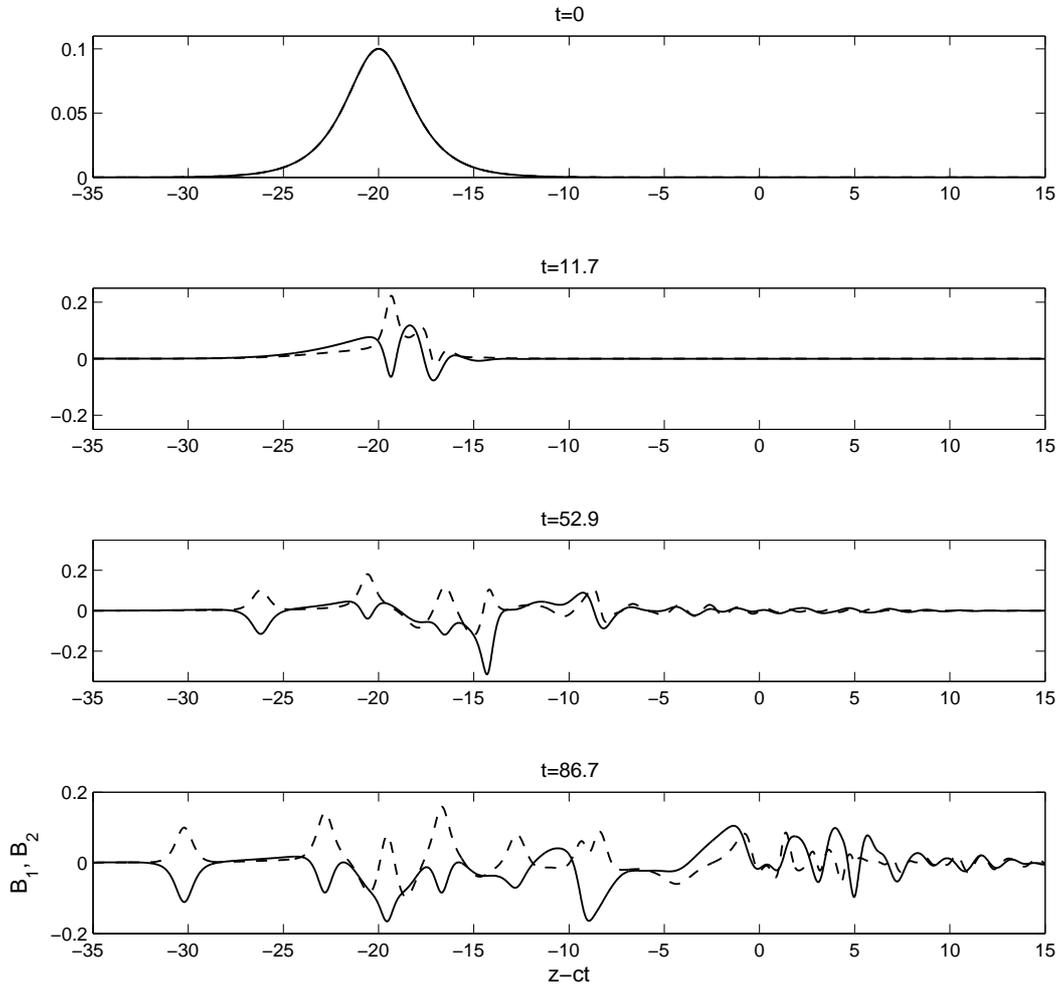}
  \caption{\small\em Evolution of a perturbation under the viscous CH dynamics: wave components $B_1$ (solid) and $B_2$ (dash) at different instants of times ($\RE = 8000$, speed of the reference frame $c = 0.75$).}
  \label{fig:fig10}
\end{figure}

\begin{figure}
  \centering
  \includegraphics[trim=0.8cm 0.8cm 1.6cm 0.8cm, clip=true, width=0.8\textwidth]{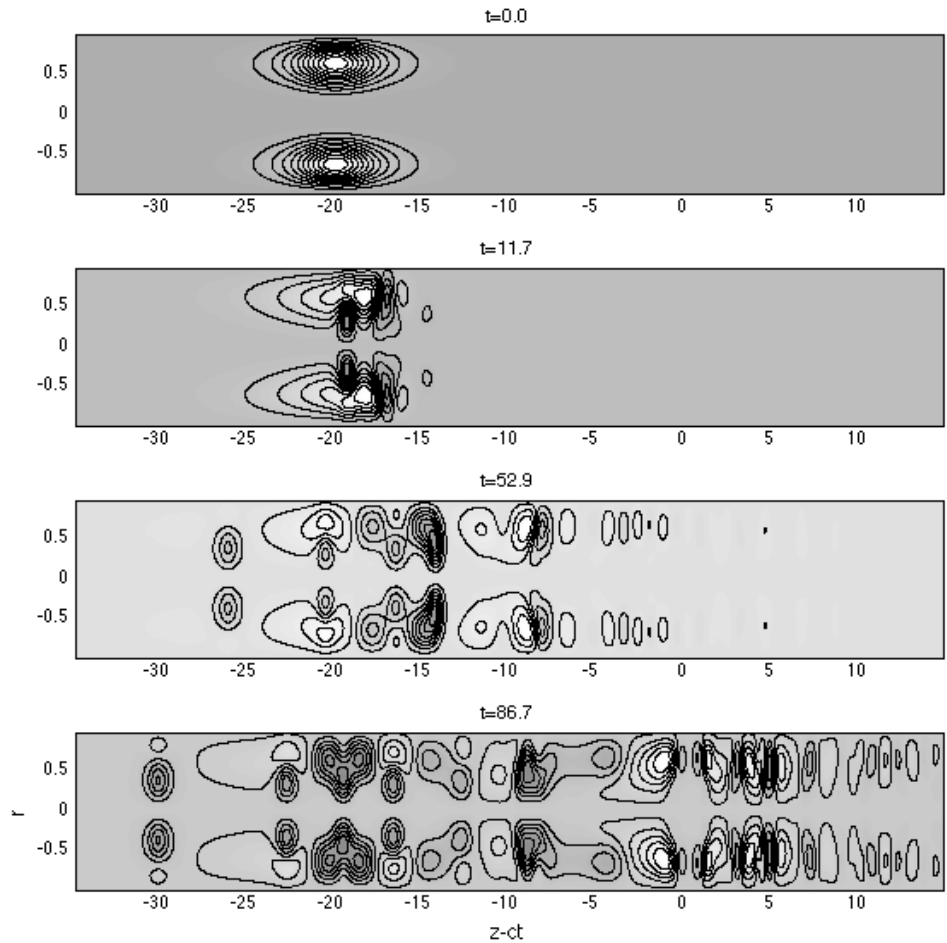}
  \caption{\small\em Long-time evolution of a perturbation under the viscous CH dynamics: streamlines of the vortical structures associated to the wave components of Fig.~\ref{fig:fig10}.}
  \label{fig:fig11}
\end{figure}

\begin{figure}
  \centering
  \includegraphics[trim=0.8cm 0.8cm 1.2cm 0.8cm, clip=true, width=0.5\textwidth]{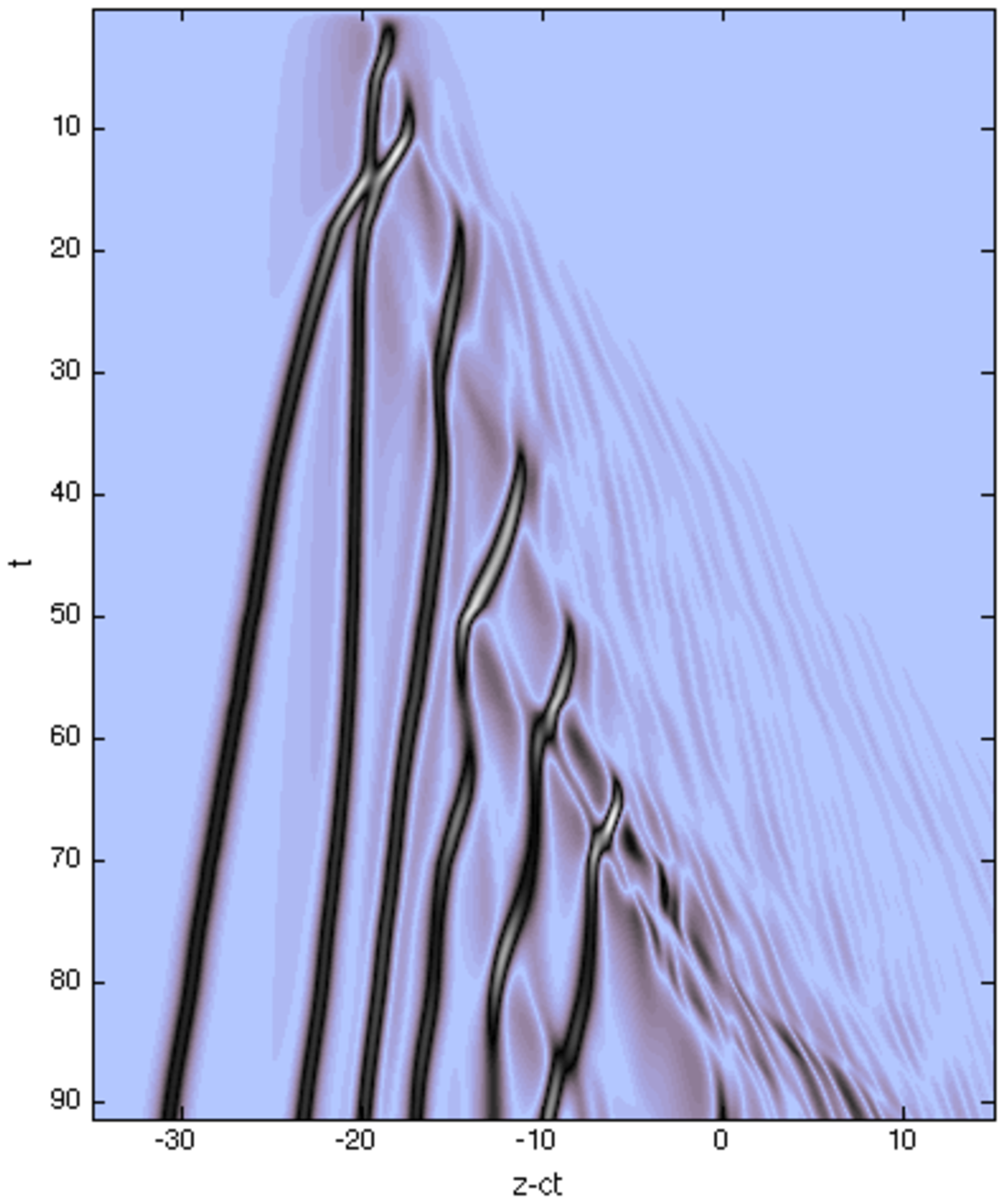}
  \caption{\small\em Space-time evolution of $\beta = |B_1 - B_2|$ for $\RE = 8000$, speed of the reference frame $c = 0.75$. Large values of $\beta$ trace centre vortexons ($B_1$ and $B_2$ have opposite sign), whereas smaller values are associated to wall vortexons ($B_1$ and $B_2$ have same sign).}
  \label{fig:fig12}
\end{figure}

Hereafter, we investigate the dynamical evolution of a localized disturbance under the two-component CH dynamics with dissipation. To do so, we exploit a highly accurate Fourier-type pseudo-spectral method to solve the CH equations (\ref{eq:CH}) as described in \citeasnoun{Fedele2012}). For $\RE = 8000$ Figure~\ref{fig:fig10} depicts snapshots of the two-component CH solution at different times and the streamlines of the associated vortical structures are shown in Figure~\ref{fig:fig11} . As time evolves, the waveform of each component steepens up and then splits into solitons and radiative waves as a result of the competition between the laminar-flow-induced wave dispersion and the nonlinear energy cascade associated to the CH nonlinearities. In physical space the initial vortical structure first compresses as a result of wave steepening and then splits into a centre vortexon and patches of vorticity in the form of wall vortexons. These may further split causing the formation of new centre and wall vortexons until viscous effects attenuate them and annihilate splitting on the time scale $t\sim \O(\RE^{6.25})$ \cite{Fedele2012b}. The formation of a vortexon slug is clearly evident in Figure~\ref{fig:fig12}, in which we report the space-time plot of the difference $\beta = |B_1 - B_2|$ of the two wave components. Here, centre vortexons correspond to larger values of $\beta$ ($B_1$ and $B_2$ have opposite sign), whereas smaller values of are associated to wall vortexons ($B_1$ and $B_2$ have the same sign). The centre vortexon arises due to a radial flux $F_{\theta r}^{(\omega)} \simeq u\omega_\theta$ of azimuthal vorticity $\omega_\theta$ from the wall to the pipe axis. This is the mechanism of inverse cascade of cross-stream vorticity in channel flows identified by \citeasnoun{Eyink2008}. Similar dynamics is also observed for long-wave disturbances associated to the KdV equations (\ref{eq:KdV}) \cite{Fedele2012c}.

Note that a vortexon slug is similar to the spreading of puffs in pipe turbulence at transition \cite{Avila2011}, but they originate from different physical mechanisms. In realistic flows, a turbulent slug arises when new puffs are produced faster than their decay in the competition between puff decay (death) and puff splitting (birth) processes. Instead, a vortexon slug arises as an inviscid competition between dispersion and nonlinear steepening of radial structures that are advected in the streawise direction by the laminar flow.

Clearly, vortexon slugs are not the realistic slugs observed in experiments, which also have a non-axisymmetric component. However, similarly to the inviscid neutral modes found by \citeasnoun{Walton2011}, centre vortexons most likely are unstable to non-axisymmetric disturbances, and may persist viscous attenuation as precursors to puffs and slugs.

Finally, we note that observed vortex compression/splitting is also evident in the numerical simulations of the propagation of nonlinear Kelvin waves and fronts on the equatorial thermocline \cite{Fedorov1995,Fedorov2000}. This is expected since the geostrophic flow is two dimensional in nature and the associated dynamical equations can be reduced to KdV/CH-type models \cite{Benney1966}.

\section{Conclusions}

We have shown that the axisymmetric Navier Stokes equations for non-rotating Poiseuille pipe flows can be reduced to a set of coupled Camassa--Holm type wave equations. These support inviscid and regular traveling waves that are computed numerically using the Petviashvili method. The associated flow structures are localized toroidal vortices or vortexons that travel slightly slower than the maximum laminar flow speed, in agreement with the theoretical predictions by \citeasnoun{Fedele2012b}. The vortical disturbance can be localized near the wall (wall vortexon) or wrap around the pipe axis (centre vortexon). Moreover, we also discovered numerically special traveling waves with wedge-type singularities, \emph{viz}. peakons, which bifurcate from smooth solitary waves. In physical space they correspond to localized toroidal vortical structures with discontinuous radial velocities (singular vortexon). The existence of such singular solutions is confirmed by an analytical solution of exponentially shaped peakons of the uncoupled wave equations. Clearly, the inviscid singular vortexon could be an artefact of the Galerkin truncation of the axisymmetric Euler equations that are projected onto the function space spanned by the first few Stokes eigenmodes. Viscous dissipation rules out the existence of peakons and the Camassa--Holm dynamics involves only regular vortexons. Indeed, we found numerically that an initial perturbation evolves into a vortexon slug, \emph{viz}. a solitonic sea state of centre vortexons that split from patches of near-wall vorticity due to an inverse radial flux of azimuthal vorticity from the wall to the pipe axis in agreement with the cross-stream vorticity cascade of \citeasnoun{Eyink2008}.

Finally, we wish to emphasize the relevance of this work to the understanding of transition to turbulence. For chaotic dynamical systems the periodic orbit theory (POT) in \cite{Cvitanovic1991} and \cite{Cvitanovic1995} interpret the turbulent motion as an effective random walk in state space where chaotic (turbulent) trajectories visit the neighborhoods of equilibria, travelling waves, or periodic orbits of the NS equations, jumping from one saddle to the other through their stable and unstable manifolds \cite{Wedin2004,Kerswell2005,Gibson2008}. Non-rotating axisymmetric pipe flows do not exibhit chaotic behaviour (see, e.g., \cite{Patera2006,Willis2008a}), and so the associated KdV or CH equations (even with dissipation). However, forced and damped KdV/CH equations are chaotic and the attractor is of finite dimension (see, for example, \cite{Cox1986,Grimshaw1994}). Thus, the study of the reduced KdV-CH equations associated to forced axisymmetric Navier--Stokes equations using POT may provide new insights into understanding the nature of slug flows and their formation.

\section*{Acknowledgements}

F.~\textsc{Fedele} acknowledges the travel support received by the Geophysical Fluid Dynamics (GFD) Program to attend part of the summer school on ``\textit{Spatially Localized Structures: Theory and Applications}'' at the Woods Hole Oceanographic Institution in August 2012. D.~\textsc{Dutykh} acknowledges the support from ERC under the research project ERC-2011-AdG 290562-MULTIWAVE.

\appendix

\section{Coefficients in Camassa--Holm equations}\label{app:a}

\begin{equation*}
  c_{jm} = -\int\limits_0^1 W_{0}\phi_{j}\L\phi_{m}\;r^{-1}\,\ud r,\quad
  \alpha_{jm} = -\int\limits_0^1 \phi_j\phi_m\;r^{-1}\,\ud r, \quad
  \beta_{jm} = -\int\limits_0^1 W_{0}\phi_j\phi_m r^{-1}\,\ud r,
\end{equation*}
\begin{equation*}
  F_{jnm} = -\int\limits_0^1\phi_j\left[\partial_r\phi_n\L\phi_m - \partial_r\left(\L\phi_n\right)\phi_m + 2r^{-1}\L\phi_n\phi_m\right]r^{-2}\,\ud r,
\end{equation*}
\begin{equation*}
  H_{jnm} = -\int\limits_0^1\phi_j\phi_m\partial_r\phi_n r^{-2}\,\ud r, \quad
  G_{jnm} = -\int\limits_0^1\phi_j\left[-\phi_m\partial_r\phi_n + 2r^{-1}\phi_n\phi_m\right]r^{-2}\,\ud r.
\end{equation*}

\section{Peakons of the dispersive CH equation}\label{app:b}

To simplify the analysis, we drop the subscripts in (\ref{eq:CH2}) and consider
\begin{equation}\label{eq:ref1}
 B_{t} + \alpha B_{xxt} + cB_{x} + \beta B_{xxx} + FBB_x + GB_xB_{xx} + HAA_{xxx} = 0.
\end{equation}
The ansatz for a peakon is
\begin{equation*}
  B=\left\{\begin{array}{c}
    a\mathrm{e}^{-\gamma s(x-Vt)}, \quad s = 1, \quad x > Vt, \\
    \\
    a\mathrm{e}^{-\gamma s(x-Vt)}, \quad s = -1, \quad x < Vt,
\end{array}\right.
\end{equation*}
Substituting this into (\ref{eq:ref1}) yields
\begin{equation*}
  \mathrm{e}^{-\gamma s(x-Vt)}W_{1} + \mathrm{e}^{-2\gamma s(x-Vt)}W_{2} = 0, \qquad s = \pm 1,
\end{equation*}
where the coefficients $W_{j}$ do not depend on $s$ and are given by
\begin{equation*}
  W_{1} = -c + V - \gamma^{2}(\beta - \alpha V), \qquad 
  W_{2} = F + \gamma^{2}(G+H).
\end{equation*}
Imposing $W_{1}=0$ and $W_{2}=0$ yield
\begin{equation*}
  V = \frac{c + \beta\gamma^{2}}{1 + \alpha\gamma^{2}}, \qquad
  \gamma^{2} = -\frac{F}{G+H}.
\end{equation*}
Peakons exist if $\gamma^{2}>0$, but we still need to find their amplitude $a$. To do so, let us consider the general ansatz
\begin{equation*}
  B = R(\xi) = R(x-Vt),
\end{equation*}
where $R$ follows from (\ref{eq:ref1}) and it satisfies 
\begin{equation*}
  -VR_{\xi} - \alpha VR_{\xi\xi\xi} + cR_{\xi} + \beta R_{\xi\xi\xi} + FRR_{\xi} + GR_{\xi}R_{\xi\xi} + HRR_{\xi\xi\xi} = 0,%\label{1-1}
\end{equation*}
and subscripts denote derivatives with respect to $\xi$. This can be written as
\begin{equation}\label{eq:refr}
  \left((c-V)R + (\beta - \alpha V + HR)R_{\xi\xi} + FR^{2}/2+(G+H)R_{\xi}^{2}/2\right)_{\xi} = 0.
\end{equation}
Clearly, if a peakon exists the term $(\beta-\alpha V+HR)R_{\xi\xi}$ must vanish at $\xi=0$, or $x = Vt$, because it is the only distributional term in (\ref{eq:refr}) that yields derivatives of Dirac functions. Thus, the peakon amplitude $a = R(\xi = 0) = \frac{V\alpha - \beta}{H}$.

%%% Bibliography %%%
\bibliography{biblio}
\bibliographystyle{harvard}

\end{document}